\documentstyle[preprint,
               aps]
               {revtex}

\begin{document}

\title{Glassy properties and localization of interacting 
       electrons in two-dimensional systems}

\author{D. Menashe$^{1}$, 
    	O. Biham$^{1}$,
	B. D. Laikhtman$^{1}$, and
        A. L. Efros$^{2}$}

\address{$^{1}$ Racah Institute of Physics, Hebrew University,
Jerusalem 91904, Israel}

\address{$^{2}$ Department of Physics, University of Utah,
Salt Lake City, Utah 84112}

\maketitle



\begin{abstract}
We present a computer simulation study of a disordered 
two-dimensional system of localized interacting electrons
at thermal equilibrium.
It is shown that the configuration of occupied sites
within the Coulomb gap
persistently changes at temperatures 
much less than the gap width. 
This is accompanied by 
large time dependent fluctuations of the
site energies. 
The observed thermal equilibration at low temperatures
suggests a possible 
glass transition only at $T=0$.
We interpret the strong fluctuations in the occupation numbers
and site energies
in terms of the drift of the system between 
multiple energy minima. 
The results also imply that
interacting electrons may be effectively delocalized 
within the Coulomb gap.
Insulating properties, such as hopping conduction, appear 
as a result of long 
equilibration times associated with glassy dynamics.
This may shine new light on the relation between
the metal-insulator transition and glassy behavior.\\ \\ 
pacs: 72.20.Ee,75.10.Nr,64.70.Pf
\end{abstract}

\newpage


The existence of glassy properties in a system of strongly
disordered localized interacting electrons was predicted
some time ago
\cite{lee}, thus introducing the terms
``Coulomb glass'', or
``Electron glass''.
Recently, experimental confirmation of this concept has
been obtained through the observation of
very slow relaxation
times characteristic of glassy dynamics \cite{zvi,zvi_new}.
Another important manifestation of glassy properties, namely
the existence of multiple low energy minima
in the energy landscape of the system,
has been known since the first computer
simulations of the Coulomb glass
\cite{bar}.
However, while there is strong
evidence for glassy behavior, no
finite temperature thermodynamic glass transition has been
observed either experimentally or numerically.
Furthermore, the two dimensional (2D) Coulomb glass has
much in common with various 2D spin glass (SG) models,
where there is strong numerical
evidence that no finite temperature phase transition occurs
\cite{Parisi}.
We wish to point out that our results
also support the absence of any
finite temperature phase transition in the 2D Coulomb glass.

The goal of this Letter is to study whether and how 
the multi-minima structure of the energy landscape, 
which results from electron-electron interaction,
affects the fluctuation properties of the Coulomb glass.
Previous finite temperature computer simulations have 
confirmed the
existence of a robust and stable gap in the density
of states (DS) around the Fermi level, known as the
Coulomb gap.
However, as far as we know, the following question has not
been studied:
Is the configuration of occupied sites within the gap
time independent, or does it change persistently with time
due to equilibrium fluctuations?

Such fluctuations may be very important for transport if
the equilibration time of the system is not too long.
For example, near a metal non-metal transition the 
equilibration time decreases sharply
\cite{zvi_new}.
It should be remembered that the standard theory of hopping
conduction in localized systems is based upon
percolation theory, and it assumes that the basic electronic
configurations and site energies near the Fermi level are
{\sl time independent}. This means that the percolation paths 
are also
time independent. 
It has long been suggested that this single 
particle picture of transport may be
altered by electron-electron interactions
\cite{pollak},
however, as yet no alternate picture has emerged.
For clean systems and systems
with weak disorder, it was recently 
demonstrated \cite{efr}, that the persistent
change of the occupancy configurations may
cause the
mechanism of the conductivity to change from percolation 
to diffusion.

We present here computer simulations designed to study the
effect of thermodynamic fluctuations on
the site occupation numbers and energies in the Coulomb glass.
Our results indeed show a persistent change of the occupation
configuration within the Coulomb gap,
even at temperature well below the gap width.
This persistent change creates a time dependent random
potential, which causes fluctuations of the site energies.
The fluctuations are much larger than the temperature, and
for sites within the gap, 
are of the order of the gap width.
We interpret these results in terms of a drift of 
the total energy of
the system between
different minima of the energy landscape.
It can also be considered as some kind of 
classical delocalization effect.

For the purpose of our study we use the standard 
Coulomb glass Hamiltonian 
given by
\begin{equation}
H = \sum_{i}\phi_{i} n_{i} + {1 \over 2} \sum_{i \neq j}
    {e^2 \over {r_{ij}}} \left(n_i - \nu \right)
                         \left(n_j - \nu \right). 
\label{Ham}
\end{equation}
\noindent 
The electrons occupy sites on a 2D lattice, 
$n_i  = 0, 1$ 
are the occupation numbers of these sites and
$r_{ij}$ is the distance between sites $i$ and $j$.
The quenched random site energies $\phi_i$ are distributed 
uniformly within an interval $[-A, A]$.
To make the system neutral each site has a positive 
background charge $\nu e$, where  $\nu$  is  
the  average occupation number, i.e. the filling factor of the 
lattice.
We concentrate here on the 
case  $\nu=1/2$, where the Fermi level is zero 
due to electron hole symmetry. 
It is expected that the features of this model
relevant for our purpose are 
dimensionality, Coulomb interactions, 
and strong diagonal disorder. 
Hereafter we take the lattice constant 
$a$ as a length unit and 
$e^2/a$ as an energy unit.
Then, the single particle energy at site 
$i$ is given by
\begin{equation}
\epsilon _i = \phi_{i}  +  \sum_{ j}{1 \over {r_{ij}}} 
\left(n_j - \nu \right).
\label{en}
\end{equation}
It has long been established that near the Fermi level the 
long range Coulomb interaction cannot be considered as
a perturbation.
This results in a soft gap in the density of
single-particle states (DS), known as the Coulomb gap.

In this letter we 
study numerically the equilibrium fluctuations of the 
occupation numbers and site energies within the gap,
at temperature well below the gap width.
To this end
we use the standard Metropolis algorithm, where the 
rate of a hopping transition depends only on the 
energy difference between the initial and final configurations.
Specifically, the rate does not depend on the hopping 
distance, which is limited only by the system size. 
The use of such transition rates greatly decreases the
equilibration time of the system
compared to short range hopping transitions.
Note that 
the Hamiltonian of Eq.(\ref{Ham}) 
does not contain any dynamics in itself, 
and therefore the simulation time does
not reflect any physical time. 
However, averaging over the simulation time
is equivalent to 
ensemble averaging, assuming the system is in   
thermal equilibrium. 
Related approaches are 
commonly used to study the thermodynamics of 
various SG models
\cite{binder}.

The simulations were performed on a square lattice of
$L \times L$ sites with periodic boundary conditions. 
In this torus geometry, the distance between two sites
is taken as the length of the shortest path between them. 
All results obtained were averaged over $P$ different 
sets of the random energies 
$\{\phi_i\}$. Unless stated otherwise,
the values $L=50$ and 
$P=100$ were used throughout. 
Also, we use the notation whereby $\langle x \rangle$ means
time averaging of the quantity $x$,
whereas $\overline{x}$
denotes averaging over
sets of random energies $\{\phi_i\}$. 

The dynamics within the Coulomb gap can be seen in two 
ways. One is the time dependence of the single 
particle energies, which we call spectral diffusion. 
The other is the time dependence of the configuration of 
occupied sites within the gap. 
These two phenomena are closely related 
since the average occupation number of all 
sites with energy $\epsilon$
is given in thermal equilibrium by the Fermi function
\cite{efp}.
Thus, at low enough temperatures, the occupation
number of a site changes when it's energy crosses the Fermi 
level.

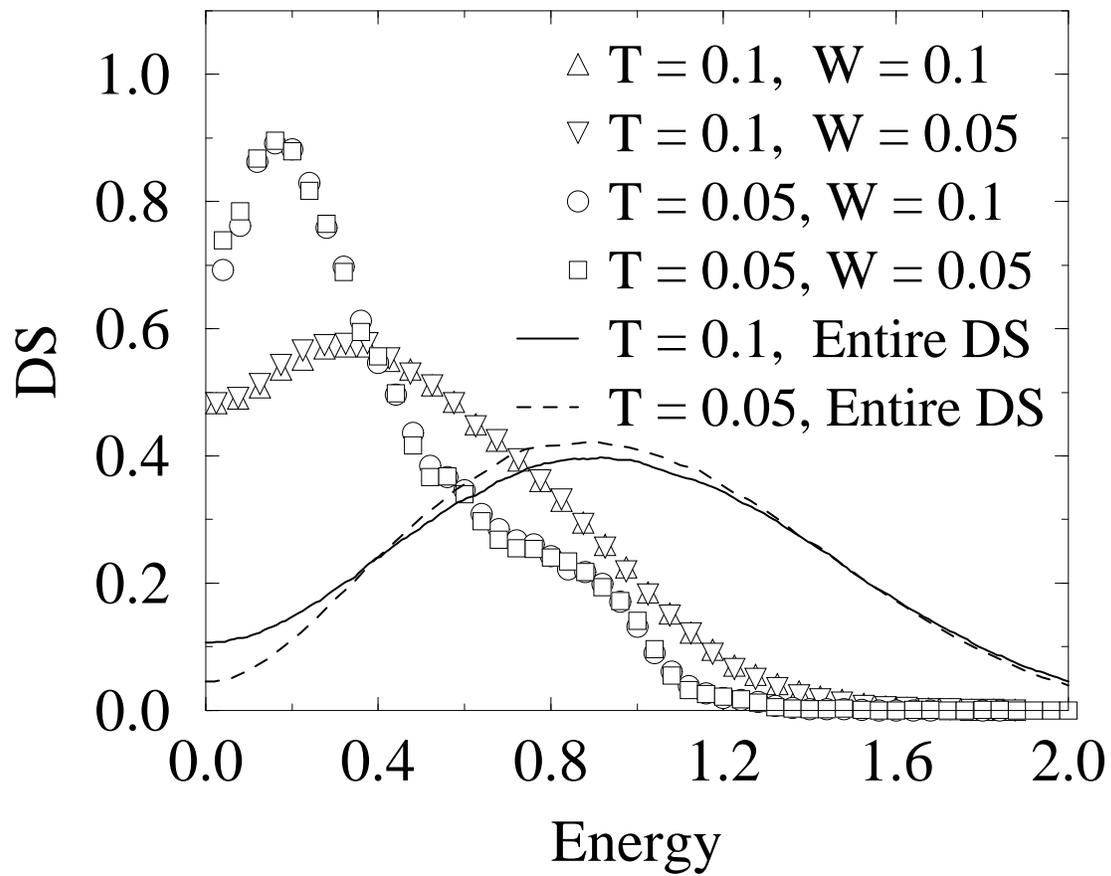
\begin{figure}
\caption{Final energy distribution of sites initially 
in the energy range $[-W,W]$. Results are shown for $A=1$ 
and different $T$ and $W$. Only positive energies are shown
due to electron-hole symmetry. 
The DS of the entire system is also shown for different $T$.}
\label{Test_Site_Dist}
\end{figure}

To study the spectral diffusion, we mark after $t_w$ 
Monte Carlo (MC) steps
all sites whose single particle energies 
are in a narrow interval $[-W,W]$ within the gap, 
and then observe the evolution of the distribution of 
these energies as the simulation proceeds. 
We find that after some number of MC steps, this
distribution reaches an asymptotic form 
which is independent of $t_w$, unless $t_w$ is shorter than
the equilibration time. 
Fig. \ref{Test_Site_Dist} 
shows the final energy distribution,
averaged over sets of $\{\phi_i\}$,
for $A=1$ and various values of the width $W$ 
and the temperature $T$. 
The DS of the entire system, 
which exhibits the Coulomb gap, is also shown.
One can see that for any given 
$T$, the final distribution does not depend on $W$ for small
$W$, and it's width is much larger than $W$. 
Furthermore, the width of the distribution is much 
larger than the temperature.

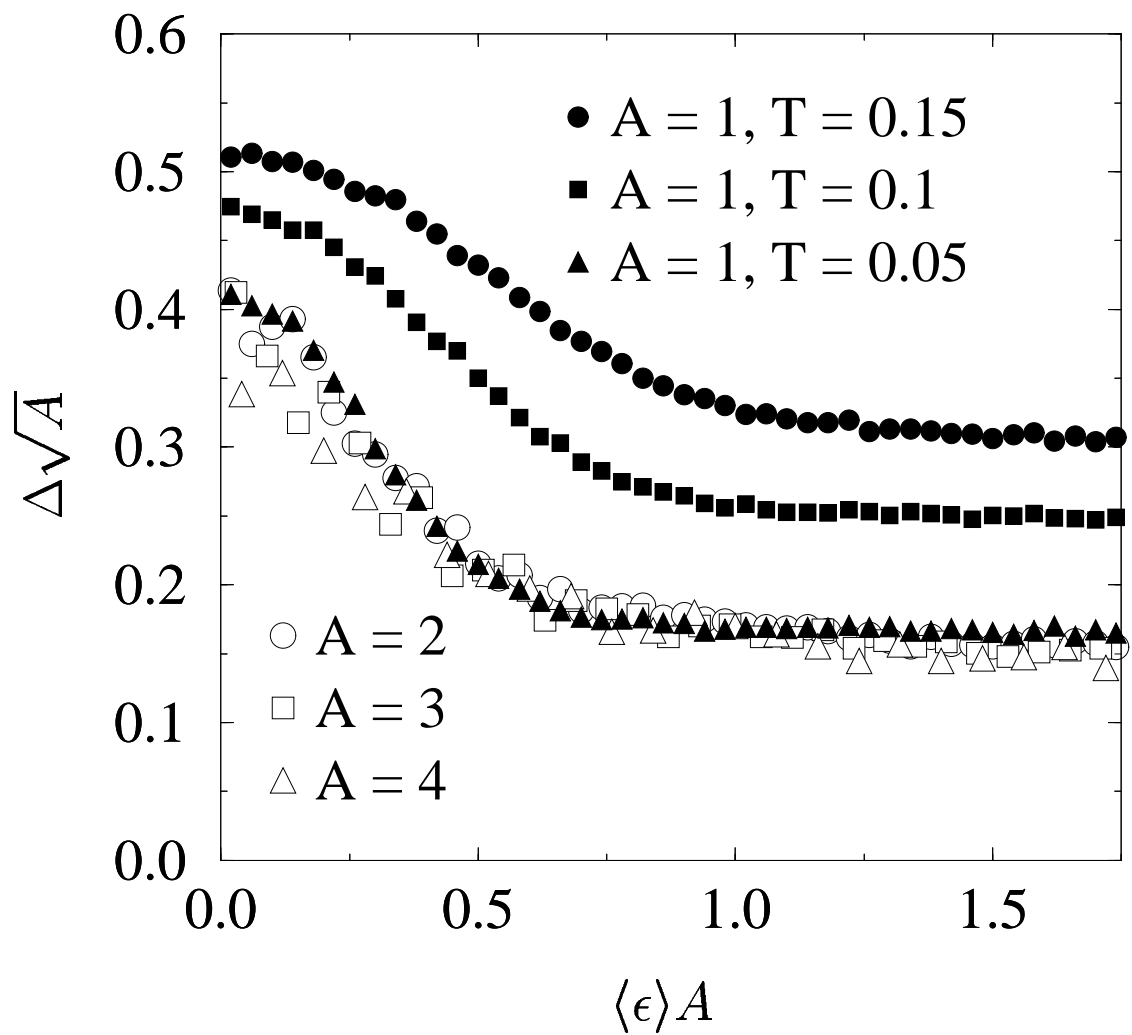
\begin{figure}
\caption{Site energy standard deviation as a
function of site average energy 
$\overline{\Delta \left(\langle\epsilon\rangle\right)}$,
for various values of $A$ and $T$. 
For $A>1$ the temperature is given by $T = 0.05/A$.}
\label{EnWidth_AvgEn}
\end{figure}

Another way to observe spectral diffusion is to measure 
the time average of the single-particle energy at 
site $i$, $\left \langle \epsilon_i \right \rangle$, 
and the standard deviation at the same site, 
$\Delta_i=\sqrt{\left \langle \epsilon_i^2 \right \rangle -
                \left \langle \epsilon_i   \right \rangle ^2}$.
We perform this calculation for all sites and create a 
function 
$\overline{\Delta(\langle\epsilon\rangle)}$. 
Fig. \ref{EnWidth_AvgEn} 
shows this function for $A=1$ and 
various temperatures. From this figure it is clear that the 
standard deviation for all sites is much larger than the 
temperature, while for sites near the Fermi level the standard
deviation is $2-3$ times larger than for other sites. 
Sites with large ${\Delta}$ are expected to be
``active'' sites,
meaning they often change their occupation
numbers as their energies cross the Fermi level.
The occupation number changes of these sites are accompanied 
by a reorganization of the local configuration of occupied 
sites, which in turn is responsible for the larger
value of ${\Delta}$.
On the other hand, the sites with smaller
${\Delta}$ are 
``passive'' sites, and change their energies only 
in response to the random time dependent potential 
created by the active sites. 

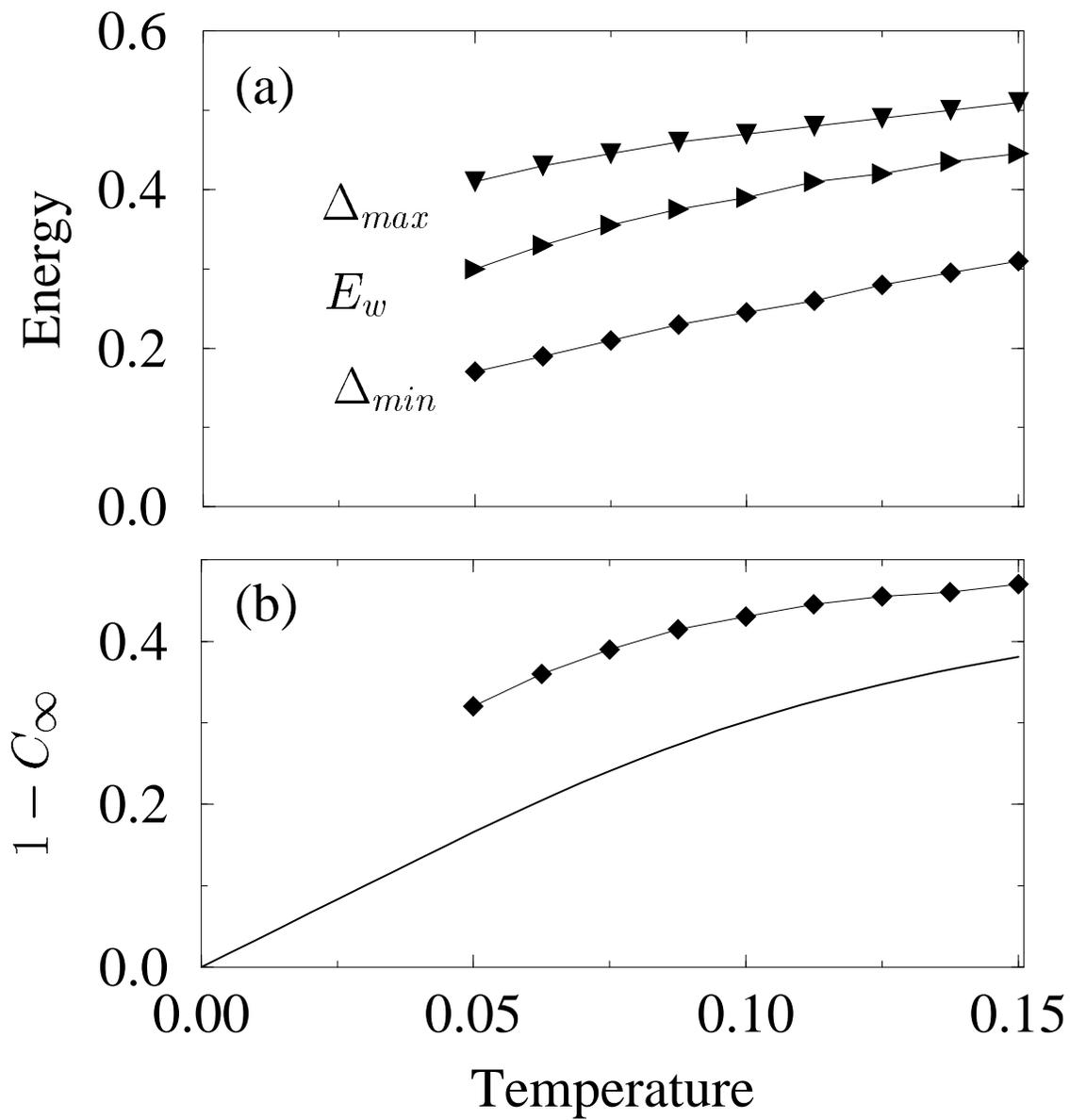
\begin{figure}
\caption{(a) Maximum and minimum values of 
$\overline{\Delta \left(\langle\epsilon\rangle\right)}$
($\Delta_{max}$
and $\Delta_{min}$), 
and the width $E_w$
as a function of temperature for $A=1$.
(b) The correlation function $1-C_\infty$ 
as a function 
of temperature for $A=1$ and $W = 0.3$. The solid
line shows the correlation function for the 
non-interacting system.}
\label{EnWidth_Temp}
\end{figure}

Fig. \ref{EnWidth_Temp}(a) 
shows the maximum and minimum values of 
the standard deviation 
$\overline{\Delta(\langle\epsilon\rangle)}$,
as a function of temperature. 
From the results it appears that 
these functions tend to a finite values 
as $T\rightarrow 0$.
It is also interesting to consider the
width of the curves in 
Fig. \ref{EnWidth_AvgEn}
as a function of temperature. 
We define this width, $E_w$, as the energy at which 
$\overline{\Delta(\langle\epsilon\rangle)}
 = \langle\epsilon\rangle$. 
The meaning of $E_w$ is that sites which satisfy 
$\langle \epsilon \rangle < E_w$ 
have energy fluctuations 
larger then their average energy, and therefore
are active. 
From Fig. \ref{EnWidth_AvgEn} 
it is also apparent that these sites 
have larger value of ${\Delta}$, thus
supporting our understanding that these are indeed the
active sites of the system. 
The width $E_w$ as a function of temperature is plotted in 
Fig. \ref{EnWidth_Temp}(a), 
and it also appears to tend to 
a finite value as $T\rightarrow 0$.

The above results may indicate
that the active sites
are predominantly within the Coulomb gap. 
This is reasonable, since the occupation 
number of sites within the gap
is strongly affected by interactions.  
However, at $A=1$ all characteristic energies, including
the gap width, are of order unity. 
Thus, to check whether 
the active sites are indeed within the 
gap, it is necessary to simulate $A >1$. 
Then, the width of the gap decreases with $A$
as $E_g \sim 1/A$. The results from these simulations are also
presented in 
Fig. \ref{EnWidth_AvgEn}, where 
$\overline{\Delta(\langle\epsilon\rangle)} \sqrt{A}$ 
is plotted as  
a function of $ \langle\epsilon \rangle A$.
For these plots we use $L = 200$ and $P = 20$, and 
the temperature for each $A$ is $T=0.05/A$, keeping
it constant in units of the gap width.
Using this scaling, the curves for 
$A>1$ collapse into one, and correspond to the curve for 
$A = 1, T = 0.05$.  
Thus, the energy region containing active 
sites scales as $1/A$, and the active sites are indeed
within the Coulomb gap.
Moreover, since the number of active sites decreases with
increasing $A$, the values of $\Delta$ should also 
decrease.  This is indeed supported by the data in 
Fig. \ref{EnWidth_AvgEn}. 

Thus, it appears that the configuration of occupied sites 
within the Coulomb gap persistently changes in 
thermodynamic equilibrium. To obtain
more information about this motion, 
one can study the correlation function of occupation 
numbers.
We do this by constructing
a vector ${\bf D}(t_w)$ after
$t_w$ MC steps have been performed, 
whose components are the occupation numbers
$n_i$ of all 
sites within a given energy range $[-W,W]$. The vector is
normalized so that  
${\bf D}(t_w) \cdot {\bf  D}(t_w)=1$.
As the simulation proceeds, we check the 
occupation number of these same sites, construct the vector
${\bf D}(t_w + t)$, and 
calculate  
$C(t_w, t)= \overline {{\bf D}(t_w){\bf \cdot D}(t_w+ t)}$.
Correlation functions analogous to $C(t_w, t)$ 
are commonly used to measure the similarity 
between two configurations
\cite{Parisi}. 
For identicle configurations
$C(t_w, t)=1$,
while if there is no correlation
$C(t_w, t)=0.5$. 
Basically, we are interested in
$C_\infty = \lim_{t_w \rightarrow \infty}
               \lim_{t \rightarrow \infty} 
               C(t_w, t)$, 
which is a measure of the similarity of two arbitrary 
states of the system at thermal equilibrium. 
For a non-interacting system, 

\begin{equation}
	C_\infty={\int_{-W}^W f^2(\phi)d\phi
	\over \int_{-W}^W f(\phi)d\phi},
\label{s}
\end{equation}
where $f(\phi)$ is the Fermi function.
Thus, for the non-interacting system  $C_\infty = 1-T/W$
at $W\gg T$  and $C_\infty = 0.5$ at $T\gg W$. 

In order to evaluate $C_{\infty}$ from the simulation, 
we measure $C(t_w, t)$ as a function of $t$ 
for a given $t_w$, and wait long enough so that 
$C(t_w,t)$ becomes 
independent of $t$. We denote this saturated value 
as $C(t_w,\infty)$.
We then increase $t_w$ until $C(t_w,\infty)$ becomes 
independent of $t_w$, and thus obtain our estimate of
$C_\infty$. 
The value of $t_w$ at which $C(t_w,\infty)$ becomes 
independent of $t_w$, is the equilibration 
time $\tau_{eq}$ of the system. 

In this light, an important question is whether the Hamiltonian 
of Eq. (\ref{Ham}) 
exhibits a finite temperature glass 
transition in 2D.
If so, then below the transition temperature the 
equilibration 
time should increase with system size $L$, 
and our results may also depend on $L$.
To test this we have studied
the size dependence of the equilibration times
\cite{els}, 
and found that within
the temperature range studied here, namely $T \geq 0.05$,
the equilibration times (measured in MC steps/site) saturate as
a function of $L$. The value of $L$ at which the equilibration
times become $L$-independent is the correlation 
length of the system, and is smaller than the system 
size we use to produce the results. 
The $L$-independent equilibration times strongly 
increases with decreasing temperature, 
making it difficult to study temperatures below $T = 0.05$.
However, since at $A=1$ the temperature $T=0.05$ 
is well below any 
relevant energy scale, 
we conclude that there is no finite temperature 
phase transition. 

The results for $C_\infty$ for the interacting system 
are shown in 
Fig. \ref{EnWidth_Temp}(b)
as a function of temperature,
for $A=1$ and  $W=0.3$.
The corresponding function for the non-interacting system,
calculated directly from 
Eq. (\ref{s}), 
is also shown.
Attempting to extrapolate the results to 
lower temperature, 
we obtain 
$\lim_{T \rightarrow 0}(1-C_\infty)
\approx 0.15$.
Thus, for any two different configurations
in thermal equilibrium 
as $T$ approaches zero, 
about $15\%$ 
of the sites within the energy interval $[-0.3,0.3]$ will 
have different occupation numbers. 
This also means that about $30\%$
of sites in this interval 
are active.
Note that by increasing $W$ we include more sites in 
the correlation function $C_\infty$, but according to 
our results for the spectral diffusion, most of these
sites remain passive as $T \rightarrow 0$. 
Thus, $C_\infty$ should increase as $W$ 
increases, as we indeed observe for $W=0.6$. 

We view the large values of 
$1-C_\infty$ at low $T$
as a manifestation of the
multiple minima of 
the total energy landscape of the system. 
These minima, known as pseudoground states 
(PS's) in the context of the Coulomb glass 
\cite{bar}, 
are close in energy to 
the ground state, but 
differ from each other by a finite fraction of the 
site occupation numbers. 
Their properties have 
been widely studied theoretically
\cite{shu},  
and they have been used to explain
\cite{all}
the experimentally 
observed long relaxation time 
\cite{zvi,zvi_new}. 
The details of our interpretation are as follows:
The Gaussian fluctuations of
the total system energy are
$T\sqrt{C_V L^2}$, where the 
specific heat $C_V$ is of the order of $T/E_g$.
At finite temperature
the system drifts through all PS's that are inside the 
fluctuation interval of the total energy.
This drift causes a persistent change of the occupation 
number configuration, and 
creates a time dependent random potential 
which is responsible 
for the spectral diffusion of the site energies. 
The picture should remain T-independent at very 
low $T$, providing the  
fluctuation interval contains many PS's. 
Since the number of PS's increases
exponentially with the volume of the system 
\cite{shu}, 
this last condition should be fulfilled
down to zero temperature in a macroscopic 
system.

The following example may clarify our interpretation. 
Suppose there is no disorder in the system, 
so that $A=0$. Then at low 
temperature the system forms a Wigner crystal,
which is two-fold degenerate 
on a square lattice at half filling. 
These two states represent our PS's.
If transitions between them are permitted, 
then all sites in the system continuously change their 
occupation number 
and the energy of each site fluctuates 
with a standard deviation of order unity. 
Note that a major difference between the Wigner crystal 
and the disordered system is that the former has
a gap in the excitation spectrum, while the latter
possesses a continuous spectrum of PS's.

It is important to point out that our results cannot 
be explained by assuming that the
excitations of the system are separated pairs of sites, 
with electrons hopping back and forth between the sites of 
each pair. This
assumption would mean that electrons 
are effectively 
localized in space. 
Since the energy density of such excitations 
is constant at low energies, meaning
the number of available excitations decreases linearly with 
temperature, one immediately obtains that
$\lim_{T \rightarrow 0}(1-C_\infty) \approx T/W$,
like in the non-interacting system.  
The same temperature dependence is obtained even if 
excitations involve a few
electrons that change their positions simultaneously
(so called many electron excitation \cite{pollak}).
In fact, any picture based upon confined separated excitations
which do not interact with each other
would mean that 
$\lim_{T \rightarrow 0}(1-C_\infty) \approx T/W$.
Since our data definitely contradicts this temperature
dependence, we conclude that such excitations
cannot explain our results. 

Thus, the results presented in this work
may indicate the existance of 
a classical delocalization effect which is related to 
the glassy properties of the system. The formal 
criterion 
for the existence of this effect is 
that  $\lim_{T \rightarrow 0}(1-C_\infty)$ is nonzero.
The limit $T \rightarrow 0$ should be taken in such a way that 
the thermodynamic fluctuations of the total energy are
larger than the energy distance between PS's. 
In a macroscopic system such a limit should 
always be possible. 
It is important, however,
that the equilibration time of the system increases 
greatly as the temperature goes to zero.
This increase should be even more pronounced 
if one takes into account 
the exponential dependence of the tunneling rate 
on the tunneling distance. 
Therefore, as the temperature 
decreases, the system will be 
frozen in phase space during times shorter then 
the equilibration time. We believe this to be the cause
of the insulating properties of the Coulomb  
glass, such as hopping conduction.
A possible confirmation of this view may be found in 
the experimental observation 
\cite{zvi_new}
that the equilibration time decreases sharply 
near the metal-insulator transition.
A similar view was also suggested by Pastor and 
Dobrosavljevic \cite{dobr}, however, they considered
a dynamic mean field theory with 
infinite range interaction, thus neglecting the 
effect of dimensionality. 

In summary, we have presented strong computational evidence 
that in a disordered 2D system of localized interacting
electrons, the
configuration of occupied sites within the 
Coulomb gap
persistently changes with time. 
This effect persists down to temperatures well below 
the Coulomb gap width, and is
accompanied by a time dependent random potential
responsible for fluctuations of the site energies 
within the gap. 
We view this as a classical delocalization effect, which 
may be suppressed at low temperature due to very long
equilibration times. 
This suggests that the transport properties of the system 
may be intimately related to glassy behavior. 
Thus, the system may be considered 
as a ``slow metal''. 
The increase of the localization radius
due to quantum mechanical overlap may transform it into a 
normal metal.
We thank Z. Ovadyahu and 
A. Vaknin for many helpful discussions. 
A.E. would also like
to acknowledge fruitful discussions with 
A. I. Larkin and B. I. Shklovskii. 
This work was supported by the US-Israel Binational Science
Foundation Grant 9800097, and by  
the Forchheimer Foundation.

\end{document}